# Probing the extreme planetary atmosphere of WASP-12b

Mark Swain, Jet Propulsion Laboratory, California Institute of Technology, 4800 Oak Grove Drive, Pasadena, CA 91109, USA

Pieter Deroo, Jet Propulsion Laboratory, California Institute of Technology, 4800 Oak Grove Drive, Pasadena, CA 91109, USA

Giovanna Tinetti, University College London, Department of Physics and Astronomy, Gower Street, London WC1E 6BT, UK

Morgan Hollis , University College London, Department of Physics and Astronomy, Gower Street, London WC1E 6BT, UK

Marcell Tessenyi, University College London, Department of Physics and Astronomy, Gower Street, London WC1E 6BT, UK

Michael Line, California Institute of Technology, Pasadena, CA 91106, USA

Hajime Kawahara, Department of Physics, Tokyo Metropolitan University, Hachioji, Tokyo 192-0397, Japan

Yuka Fujii, Department of Physics, The University of Tokyo, Tokyo 113-0033, Japan

Adam P. Showman, Department of Planetary Sciences and Lunar and Planetary Laboratory, The University of Arizona, 1629 University Blvd., Tucson, AZ 85721, USA

Sergey N. Yurchenko, University College London, Department of Physics and Astronomy, Gower Street, London WC1E 6BT, UK

**Abstract**

We report near-infrared measurements of the terminator region transmission spectrum and dayside emission spectrum of the exoplanet WASP-12b obtained using the HST WFC3 instrument. The disk-average dayside brightness temperature averages about 2900 K, peaking to 3200 K around 1.46 μm. We modeled a range of atmospheric cases for both the emission and transmission spectrum and confirm the recent finding by Crossfield et al. (2012b) that there is no evidence for C/O >1 in the atmosphere of WASP-12b. Assuming a physically plausible atmosphere, we find evidence that the presence of a number of molecules is consistent with the data, but the justification for inclusion of these opacity sources based on the Bayesian Information Criterion (BIC) is marginal. We also find the near-infrared primary eclipse light curve is consistent with small amounts of prolate distortion. As part of the calibration effort for these data, we conducted a detailed study of instrument systematics using 65 orbits of WFC3-IR grims observations. The instrument systematics are dominated by detector-related affects, which vary significantly depending on the detector readout mode. The 256×256 subarray observations of WASP-12 produced spectral measurements within 15% of the photon-noise limit using a simple calibration approach. Residual systematics are estimated to be ≤70 parts per million.

## 1. Introduction

Among the more than 700 currently confirmed exoplanets, WASP-12b stands out as exceptional. This transiting gas giant, with a mass of 1.39 $M_J$ and a radius of 1.83 $R_J$, orbits a 6300-K G star with a period of 1.09 days, resulting in an extraordinary level of insolation and, thus, extreme atmospheric heating (Hebb et al. 2009). Given the close proximity to the stellar primary, this system presents an opportunity to study a planetary atmosphere in a unique environment. The combination of the unusual nature of this system, the relatively bright stellar primary, and the system orientation, which provides both primary and secondary eclipse events, has made this target one of the more extensively observed exoplanet systems. Analysis of these observations has led to several noteworthy results that underscore the unique nature of this planet. For example, WASP-12b is inflated to an unusual degree that could imply significant internal heating (Ibgui et al. 2010). The atmosphere is likely extended, and the planet may be losing substantial mass through Roche lobe overflow (Li et al. 2010); there is also evidence supporting the presence of a magnetospheric bow shock (Llama et al. 2011). The planet has been proposed to be carbon rich, with a C/O ≥1 (Madhusudhan et al. 2011a), a condition that may reduce TiO and VO abundances (Madusadhan et al. 2011b).

Recently, Spitzer measurements were reported that show a large-amplitude thermal phase curve with a significant phase offset at 3.6 μm (Cowan et al. 2012).

High-precision, near-infrared spectroscopy has the potential to provide additional constraints that complement the exiting measurements of WASP-12b. Previous near-infrared spectroscopic observations with Hubble Space Telescope (HST) have detected molecules such as $H_2O$, $CO_2$, $CH_4$, and CO in three hot-Jovian-type planets (Swain et al. 2008, 2009a, 2009b, Tinetti et al. 2010) and produced important constraints on the atmosphere of a hot-Neptune (Pont et al. 2009) and Super Earth (Berta et al. 2011). The need for near-infrared measurements of WASP-12b has been partially addressed with ground-based photometry (Croll et al. 2011, Zhao et al. 2012) and spectroscopy (Crossfield et al. 2012a), although the precision of these observations was not sufficient to detect molecular features. Here we report high-precision, near-infrared spectroscopy measurements obtained with the HST.

## 2. Methods: Observations and Data Calibration
### 2.1 Observations

We observed the WASP-12b system using the WFC3 instrument with the G141 grism, which provides spectral coverage from 1.1 to 1.7 μm with a spectral resolution of R=300 at 1.38 μm. The observations reported here consist of two HST visits, each with five consecutive orbits, timed to measure a primary and secondary eclipse event (see Figure 1). For both events, the observations track the system light curve from pre-ingress to post-egress to provide a spectrophotometric baseline from which the eclipse depth can be measured. The primary eclipse (transit), when the planet blocks some of the light from the stellar primary, probes the transmission spectrum of the planet's terminator region atmosphere. The secondary eclipse (occultation), when the planet passes behind the stellar primary, probes the emission spectrum of the planet's dayside atmosphere. For these observations, the primary and secondary eclipse measurements were timed to be separated by the minimum possible time feasible with HST with the objective of minimizing any effects due to temporal changes in the exoplanet atmosphere or parent star. After both the primary and secondary eclipse observation sequence, we observed a calibrator star, HD 258439, with two consecutive orbits. Detector-specific configuration information, such as integration time and subarray size, is contained in Table 1. Our choices for the detector readout mode maximized the instrument efficiency and avoided the WFC3 overheads that can reduce significantly the instrument efficiency when observing relatively bright objects, such as transiting exoplanet systems.

The near-infrared spectral signatures of transiting exoplanets, for both transmission and emission measurements, require precision measurements; consequently, the potential impact of instrument systematic errors must be considered. A purpose-built instrument, which would eliminate many of these problems, does not yet exist, and the field has to make the best of the instruments available. This underscores the need for careful investigation and a thorough understanding of instrument characteristics. This is best done by a systematic analysis of a large amount of data so that characteristic instrument behavior patterns can be identified. WFC3 is relatively new and an investigation of the instrument is especially timely.

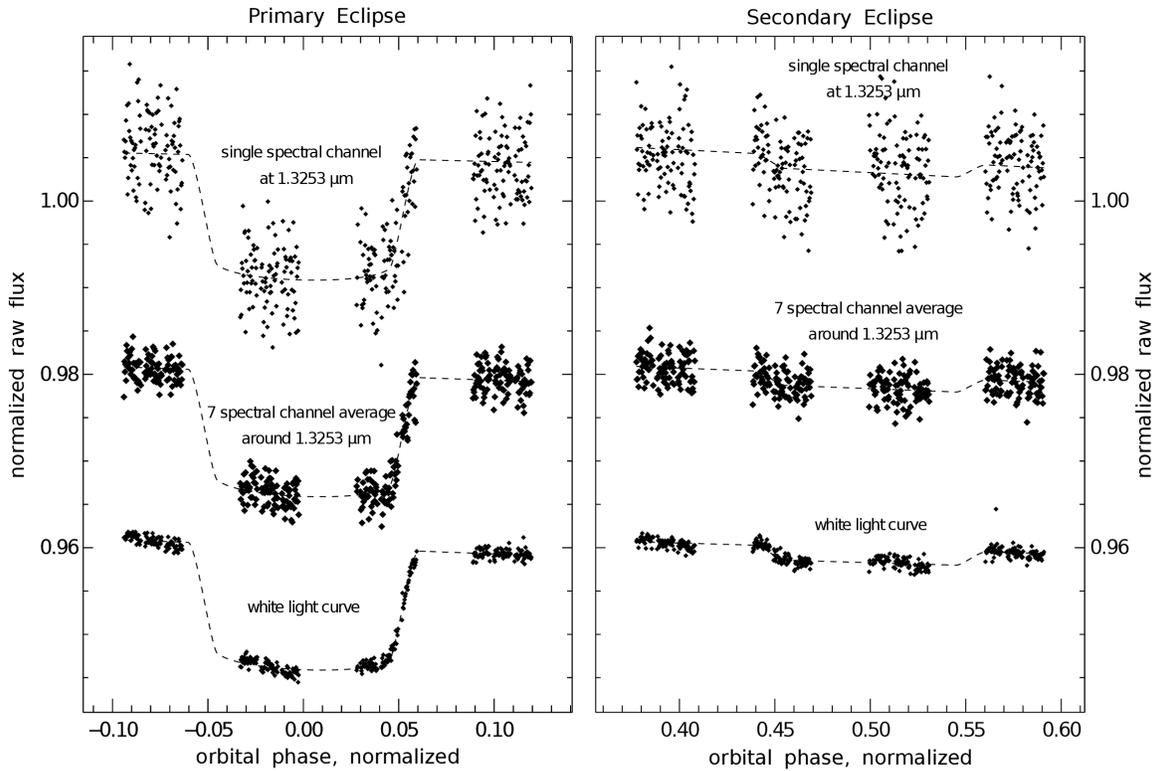

***Figure 1:*** *The WFC3 observation of the WASP-12b primary eclipse (left) and secondary eclipse (right) showing spectral and broad-band light curves as a function of orbital phase. The light curves are based on .flt data (see text for discussion) for orbits 2-5 and have been vertically offset for display clarity. A best fit linear detrending combined with a light curve model are shown (dashed); no data filtering or decorrelation has been applied. The gaps in the data are due to Earth occultations.*

## 2.2 WFC3 Systematics

Here we report on a detailed investigation of WFC3 instrument systematic errors using a large amount of archival data. Readers primarily interested in the science results can skip to Section 2.3.

To characterize the instrument, we analyzed 65 orbits of archive data covering 10 objects, using the data from individual non-destructive reads to mitigate the complicating factors of the range of target brightness and the fact that some groups appear to have over exposed their targets. We probed the detector system by analysis of both the standard pipeline output and lower level data products (both "_raw" and "_ima" archive file types). We searched for a measurement dependence on the instrument optical state by constructing estimates for focus, spectrum position, and spectrum rotation. The spectral extraction and determination of optical state parameters was performed as described in Swain et al. 2009b. In contrast to NICMOS, in which optical state parameters such as focus and spectral position produced large instrument systematics, we find that the WFC3 instrument optical state changes, while measureable, produce negligible systematic changes in the measured spectroscopic flux density. In this regard, we confirm previous findings (Berta et al. 2011). We attribute this independence of the measured flux density on instrument optical state parameters as likely due to a high level of uniformity in the instrument focal plane array. Thus, small changes in the illumination function do not produce significant changes in the measured flux density. Another area where we see a marked improvement over NICMOS is that, with WFC3, there is less "spacecraft settling" at the beginning of the observations.

Analysis of the 65 orbits, using a consistent approach, reveals the way in which the WFC3 detector system responds to light changes with time in the 512×512 and 128×128 subarray modes; these changes can occur on intra-orbit, inter-orbit, or visit-to-visit time scales (for details, see Figures 2, 3, 4 and 5 and the Supplementary Information). Importantly, these changes are only clearly detectable by analyzing the individual non-destructive reads as the detector is sampled up the ramp during an integration time. There are also significant, systematic changes in the average detector linearity relation due to over-exposure of the array (see Figure 5). Our results are consistent with the WFC3

Instrument Handbook, where the full-well saturation level of the IR channel for science observations is given as ~40,000 DN or 100,000 electrons for the default gain of 2.5 electrons per DN. Analysis of the highly saturated HD 189733b observations shows that physical full-well capacity is between ~57,000 and ~60,000, depending on the pixel.

Detector-system-related instrument systematics, if not correctly accounted for, have the potential to compromise measurement of an exoplanet spectrum. The 256×256 subarray data we analyzed suffered less from most of these problems, which implies the root cause may be connected to the array readout process and not to fundamental detector physics. However, until the root cause of changes in the detector response to light are understood, we caution that any exoplanet spectroscopy observations need to be analyzed at the level of individual non-destructive reads. Our findings summarizing the readout mode and type/extent of detector system systematic are summarized in Table 1 and Figure 2.

*Table 1: The WFC3 IR-grism exoplanet spectroscopy program objects and data sets analyzed for systematic errors using the ratio of the measurement variance to the photon noise; results are shown for the best out of eclipse orbit. We characterized the extent of the ramp effect with the same method. Detailed difference in the systematic errors, present in some data sets, are visible in the Nsamp relative flux 1.2-series figures in the Supplementary Material.*

| WFC3 IR-grism data analyzed for instrument systematics | | | | | | | | | |
|---|---|---|---|---|---|---|---|---|---|
| Observational Parameters | | | | | | | | Detector Systematics | |
| source | yy-mm-dd | Orbits | H mag. | Spectral Type | Integration time (sec) | ND reads | Subarray | best orbit | Ramp |
| WASP-12 | 11-04-11 | 5 | 10.23 | G0 | 7.35 | 3 | 256 | 1.3 | 3.7 |
| WASP-12 | 11-04-11 | 5 | 10.23 | G0 | 7.35 | 3 | 256 | 1.4 | 3.1 |
| WASP-12 | 11-04-11 | 2 | 10.23 | G0 | 7.35 | 3 | 256 | 1.3 | 3.9 |
| HD 258439 | 11-04-11 | 2 | 9.11 | A0V | 2.24 | 9 | 256 | 1.9 | 0.5 |
| HD 258439 | 11-04-11 | 2 | 9.11 | A0V | 2.24 | 9 | 256 | 2.5 | 0.0 |
| HD 258439 | 11-04-11 | 2 | 9.11 | A0V | 2.24 | 9 | 256 | 1.8 | 0.0 |
| COROT-2 | 10-10-18 | 4 | 10.44 | G7V | 21.65 | 4 | 128 | 5.8 | 4.0 |
| WASP-4 | 10-11-25 | 5 | 10.84 | G8 | 36.01 | 7 | 128 | 2.5 | 6.7 |
| WASP-4 | 10-11-25 | 5 | 10.84 | G8 | 36.01 | 7 | 128 | 3.1 | 6.6 |
| GJ1214 | 10-10-08 | 4 | 9.09 | M4 | 5.95 | 7 | 512 | 5.6 | 4.3 |
| HAT-P-7 | 10-09-29 | 5 | 9.34 | F | 7.65 | 9 | 512 | 9.2 | 8.5 |
| HAT-13 | 10-09-08 | 5 | 9.06 | G4 | 7.65 | 9 | 512 | 12.6 | 6.0 |
| TRES-2 | 10-10-09 | 4 | 9.92 | G0V | 12.75 | 15 | 512 | 8.3 | 4.1 |
| TRES-4 | 10-11-23 | 5 | 10.35 | F | 12.75 | 16 | 512 | 4.8 | 0.1 |
| HD 189733 | 10-11-10 | 5 | 5.6 | K1 | 0.23 | 2 | 128 | Highly saturated | |
| HD 189733 | 10-09-04 | 5 | 5.6 | K1 | 0.23 | 2 | 128 | Highly saturated | |

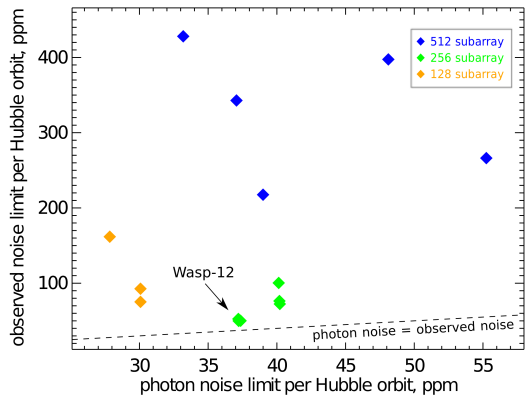

***Figure 2:*** *A comparison of detector subarray performance showing the superior performance of the 256×256 subarray. These results are based on comparing the standard deviation of the best orbit for each data set to the photon noise.*

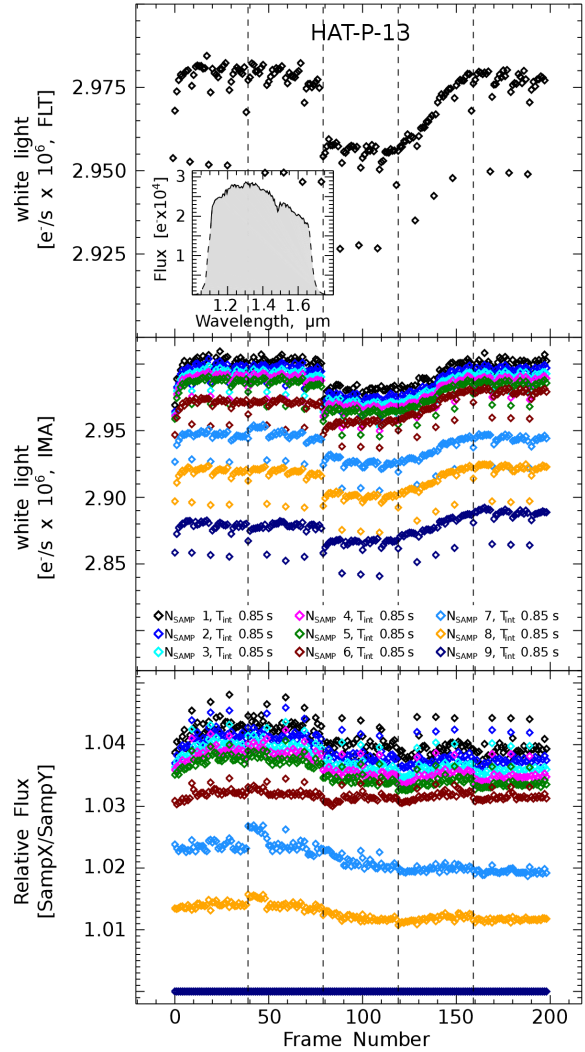

*Figure 3:* *An example of one of the standard diagnostics produced for each object in the 65 orbit data set used for assessing WFC3 systematics, with vertical dashed lines indicating a break between consecutive orbits. (Top) Broadband flux time series based on the standard pipeline data reduction. (Middle) Broadband flux time series based on individual non-destructive reads. The integration time for each non-destructive read is* $T_{int}$ *in seconds.* *(Bottom) Normalized non-destructive read time series; these provide an easy way to visualize changes in detector functionality on all time scales.*

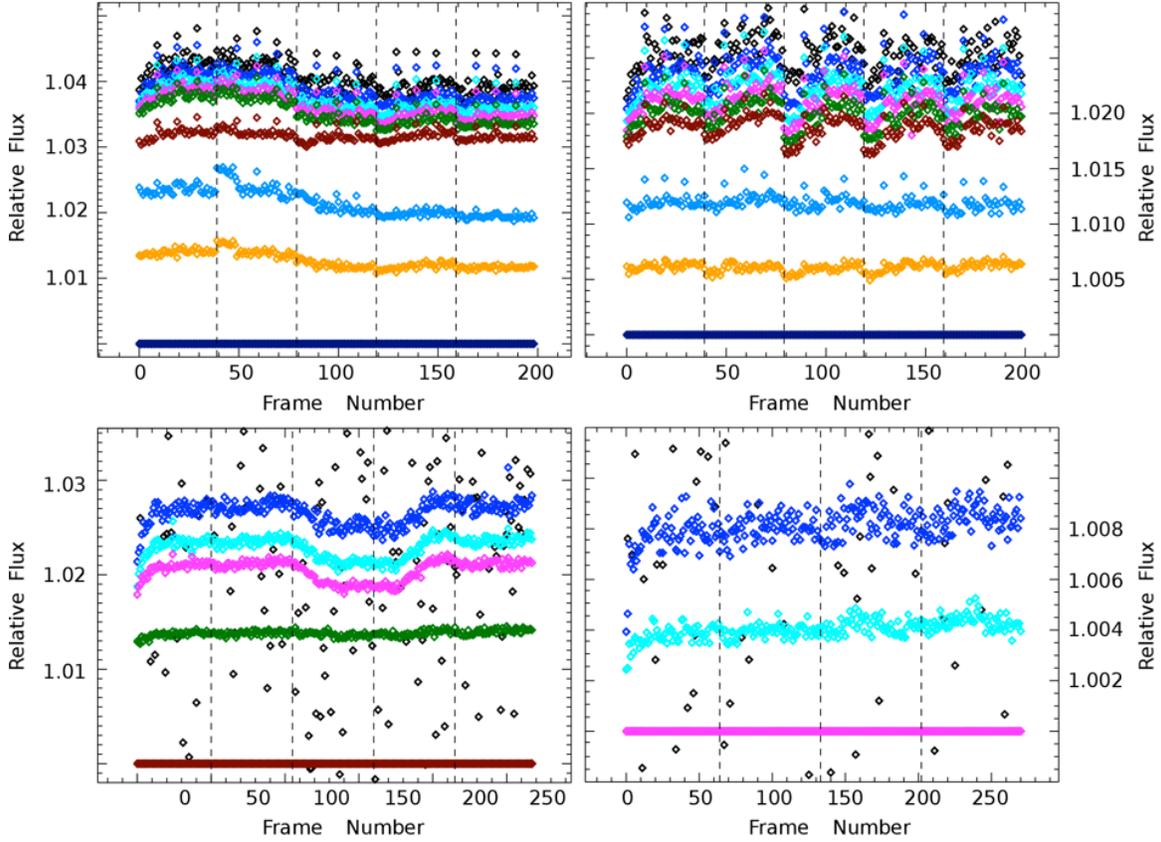

*Figure 4: Normalized Nsamp time series showing undesirable detector functionality changes (top and lower left) that could affect an exoplanet measurement. Each sample-up-the-ramp read is divided by the last read in the series. In the absence of detector-related systematics, the different time series should be constant. Instead, each of the above examples displays inter-orbit and/or intra-orbit systematics. This diagnostic is most useful in observations with several non-destructive reads (i.e., NSAMP>3). The exoplanet systems in these examples (clockwise from top left) are HAT-P-13 (512×512 subarray), HAT-P-7 (512×512 subarray), Corot-1 (128×128 subarray), and WASP-4 (128×128 subarray).*

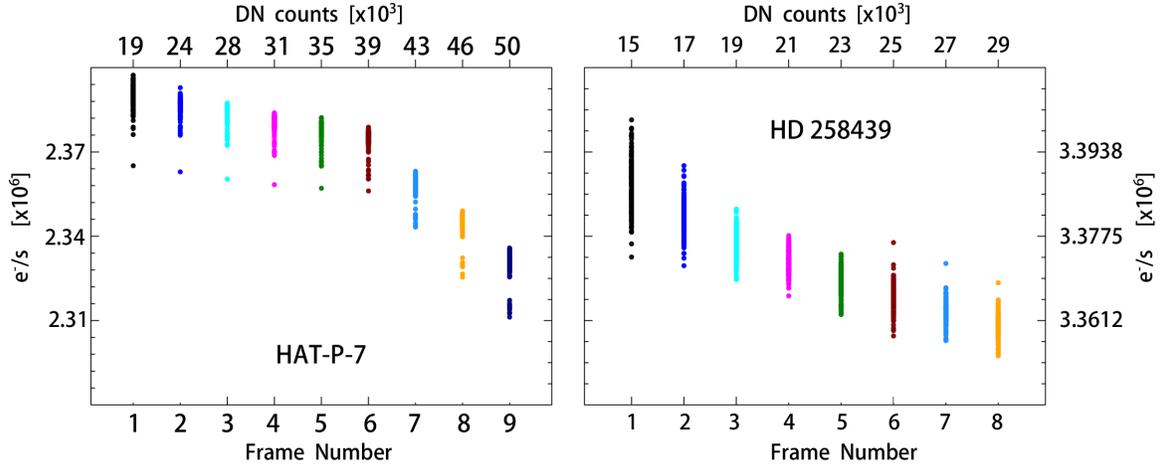

*Figure 5: The derivative of the detector linearity relation for two sources. We found a characteristic "knee" in this diagnostic when the detector well is filled beyond ~40,000 DN. A complete set of these figures is in the Supplementary information.*

142   In a recent analysis of WFC3-IR spectroscopy of the exoplanet GJ 1214b, observed in
143   the 512×512 subarray mode, Berta et al. (2011) reported a "ramp" effect of increased flux
144   measured in sequential integrations. Clarification is needed because the terminology in
145   the literature is potentially confusing; the ramp in flux measured with sequential
146   integrations reported by Berta et al. (2011) is undesirable and completely different from
147   the ramp in flux measured by consecutive nondestructive reads occurring during an
148   integration (frequently termed "sampling up the ramp"). We find that the extent of ramp
149   effect seen by Berta et al. (2011) in the flux time series is strongly correlated to the length
150   of time needed for a WFC3 buffer dump. The buffer dump operation lasts between 8 and
151   9 minutes in 512×512 subarray mode, about 3 minutes in 256×256 subarray mode, and 1
152   minute or less in 128×128 mode. In the case of the 128×128 and 256×256 subarray
153   modes, the ramp effect is minimal and, in some cases, may not be present. In contrast, the
154   512×512 subarray mode ramp feature is larger and more persistent in that it establishes a
155   trend that dominates the data for each block of measurements. During the camera buffer dump operation, the array is operated in a charge-flush mode. This charge flush mode interacts with the 512×512 subarray in a way that is qualitatively and quantitatively different from the 256×256 or 128×128 subarray modes. When Earth occultations prevent observing the source, the array is again operated in charge-flush mode. The effect of this is that there is a ramp present at the beginning of the measurements for each orbit. In the case of the 256×256 and 128×128 subarray modes, this ramp is very short, lasting only one or two samples; in the case of the 512×512 subarray modes, the ramp can continue until the next buffer dump and, for the 512×512 subarray, the effect of ~45 min. of charge flush array operation is similar to the effect of ~9 min of charge-flush operation.

For future observations of exoplanet systems with WFC3, we suggest the following guidelines.

1. Avoid the use of the 512×512 subarray mode.
2. Set the integration time so that the detector receives no more than ~40,000 DN before resetting.
3. Use a small number of samples up the ramp (for non-spatial scanning observations).
4. Work directly with the individual non-destructive read samples in the data reduction process.

Recommendation (1) avoids the intra/inter orbit variations in detector function that are seen in the 512×512 subarray mode, improves instrument overhead, and reduces the ramp effect. Recommendation (2) avoids the changes in detector function that can occur during an integration. Recommendation (3) maximizes the instrument throughput by delaying triggering a buffer dump due to the limit imposed by the file index counter.

Recommendation number (4) provides a direct way to assess any changes in detector functionality.

### 2.3  Extracting the Spectra of WASP-12b

Due to operation in the 256×256 subarray mode, WASP-12 observations are of such high quality that a simple method can be used to determine the eclipse depth. In the case of both the primary and secondary eclipse, the eclipse light curves are constructed from the Nsamp=2 .ima frame data multiplied by the exposure time of 7.35 seconds. A wavelength calibration based on band filter observations and the WFC3 dispersion relation results in an uncertainty of 5 nm/pixel. The eclipse depth in individual channels is estimated by a joint simultaneous fit to a model light curve and a linear trend in time that spans obits 2 to 5 (see Figures 6 and 7), where the slope of the linear trend and the depth parameters are the free parameters adjusted by the fitting process. The primary eclipse light curve model was generated by a 4-parameter, non-linear, limb-darkening model (Claret 2000, 2011); the limb-darkening parameters were wavelength independent (for the H-band) and were fixed during the fitting process. The system ephemeris from (Maciejewski et al. 2011) and stellar parameters were taken from Hebb et al. (2009). We find an offset of 4.7 ± 1.7 minutes with respect to the T0 of a linear ephemeris prediction (Maciejewski et al. 2011), where the standard deviation of individual spectral channels is ±0.1 minutes. Both the primary and secondary eclipse light curve models were generated using Mandel and Agol (2002) algorithms. We estimated the eclipse depth uncertainty using (1) the uncertainty as estimated by the multi-parameter (Levenberg-Marquadt type) fitting process, (2) the MCMC method, (3) a prayer-bead style residual permutation analysis, (4) and the standard deviation in the mean of the residual time series and find all methods give results consistent results for these data (e.g., for the 1.3763-μm channel, the uncertainties for these four methods are 179, 175, 178, and 170 ppm, respectively).

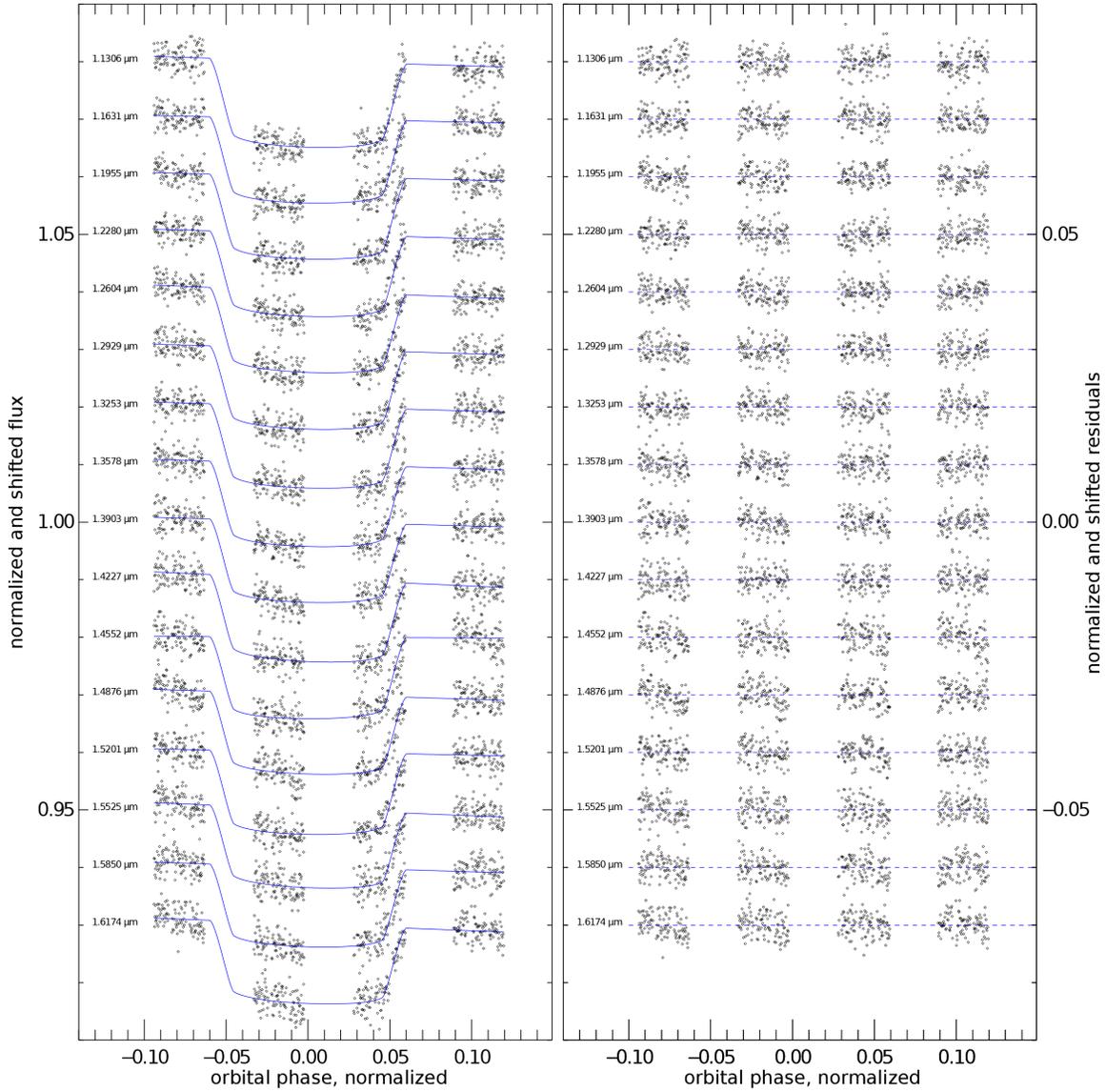

*Figure 6: The primary eclipse measurements for each spectral channel together with the best fit model shown in blue (left) and model residuals (right).*

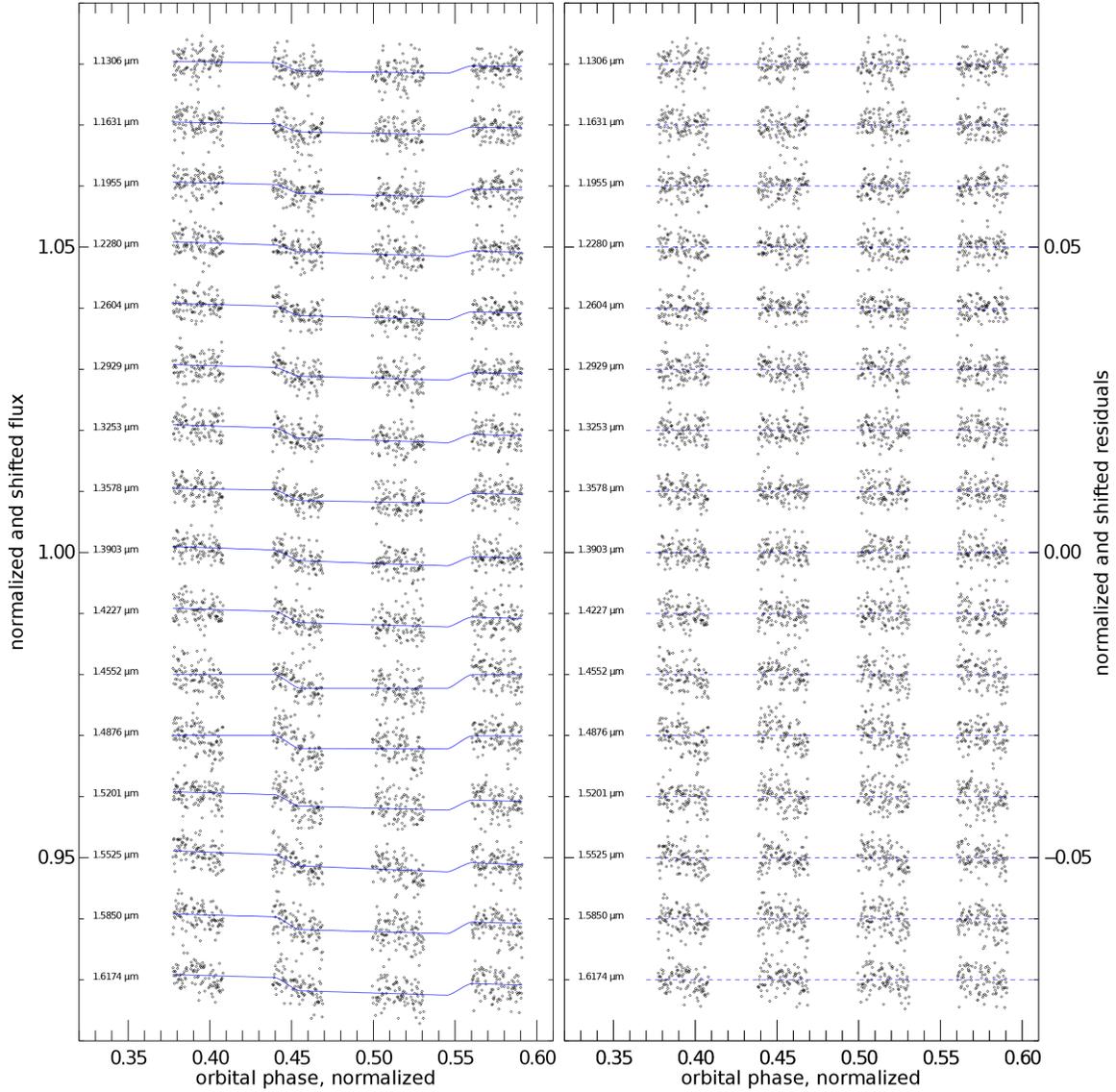

*Figure 7: The secondary eclipse measurements for each spectral channel together with the best fit model shown in blue (left) and model residuals (right).*

204    An additional correction needs to be considered in the case of WASP-12b. As noted
205 by Crossfield et al. (2012b), the presence of a nearby star, identified as Bergfors-6,
206 introduces an error in the measured eclipse depths. Bergfors-6 is visible in WFC3 narrow
207 band filter images used for our wavelength calibration and detectable as an asymmetry in
208 transverse profile of the WASP-12 system spectrum. Separation of the light from

209  Bergfors-6 from WASP-12 in our WFC3 data is possible, but requires some care. Using
210  the spectroscopy observations of HD 258439, we determined the chromatic point spread
211  function (PSF). The chromatic PSF was the used to fit two components — that of WASP-
212  12 and Bergfors-6 — for each spectral channel (see Figure 8). This allowed a
213  determination of the amount of light originating from Bergfors-6 and WASP-12b; both
214  primary and secondary eclipse depths were then corrected, following the method of
215  Kipping & Tinetti (2010), for the Bergfors-6 contamination.

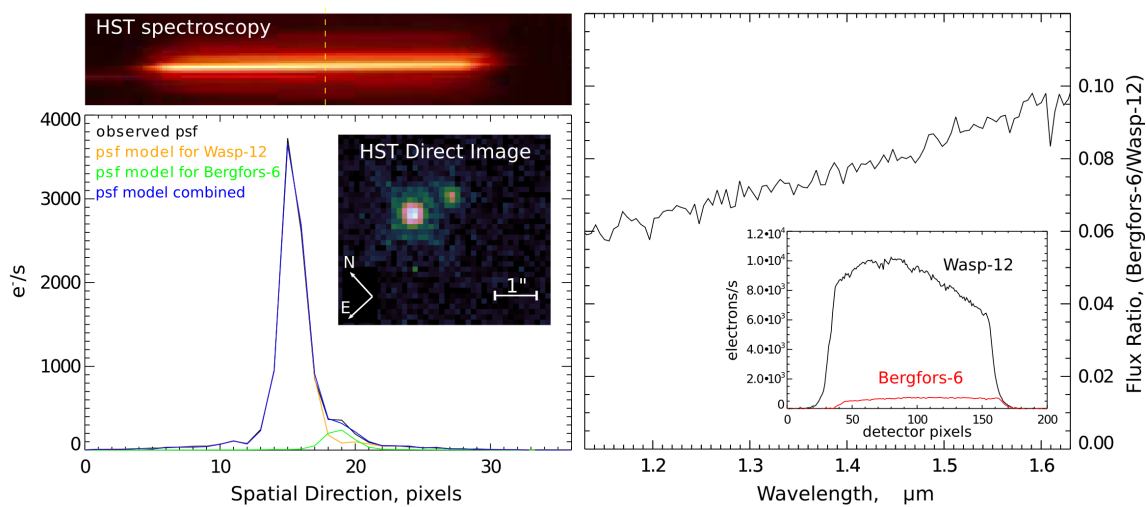

*Figure 8: Left: The WFC3 spectrum WASP-12 includes a contribution from the nearby star Bergfors-6. The WFC3 direct image of WASP-12 clearly shows the fainter object Bergfors-6 located one arcsecond to the West. The contribution from Bergfors-6 to the WASP-12 spectrum is measured by fitting two PSF components to the transverse spectral profile. Right: The measured contribution of Bergfors-6 to the measurements allowing the WASP-12 component to be isolated.*

216  To assess the quality of the linear inter-orbit detrending method, we compared the
217  measured noise to the theoretical photon noise as a function of the number of spectral
218  channels averaged to create the time series. This analysis shows that the detector behaves
219  very well in this readout mode; the presence of systemic noise becomes detectable when

220  5 or more pixels are averaged together, but the systematic noise averages (see Figure 9)
221  down as pixels to the -0.4 power (-0.5 would be ideal). The signal-to-noise of the final
222  WASP-12b spectrum is increased by averaging together seven individual channels
223  weighted by the uncertainties and achieves about 160 ppm or ~1.15 times photon noise
224  (see Figure 10 and Table 2) and avoids the kinds of questions associated with complex
225  decorrelation methods. Recently, WFC3-IR measurements of an exoplanet transmission
226  spectrum have been reported by Berta et al. (2011) using an approach termed the OOT
227  method to remove the large intra-orbit systematics found in 512×512 subarray data. As a
228  consistency check, we applied the OOT method to our WASP-12 data and find virtually
229  identical results.

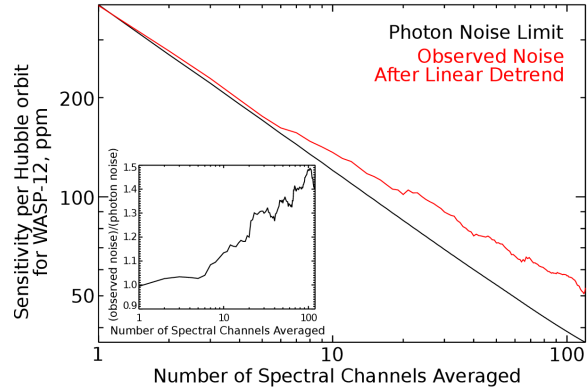

*Figure 9: The excellent performance of WFC3 is shown by a comparison of the theoretical photon-limited noise, based on detected photons, and the standard deviation of residuals for the primary eclipse measurements using a single HST orbit centered on orbital phase -0.02 (see Figure 1) as a function of the number of pixel-based spectral channels averaged together. Instrument systematics become detectable when 5 or more pixel-based spectral channels are averaged together. However, these instrument systematic errors average down very well (only slightly worse than the square root of bandwidth) and the ultimate performance of the instrument is ~1.45 times the photon noise for our observations.*

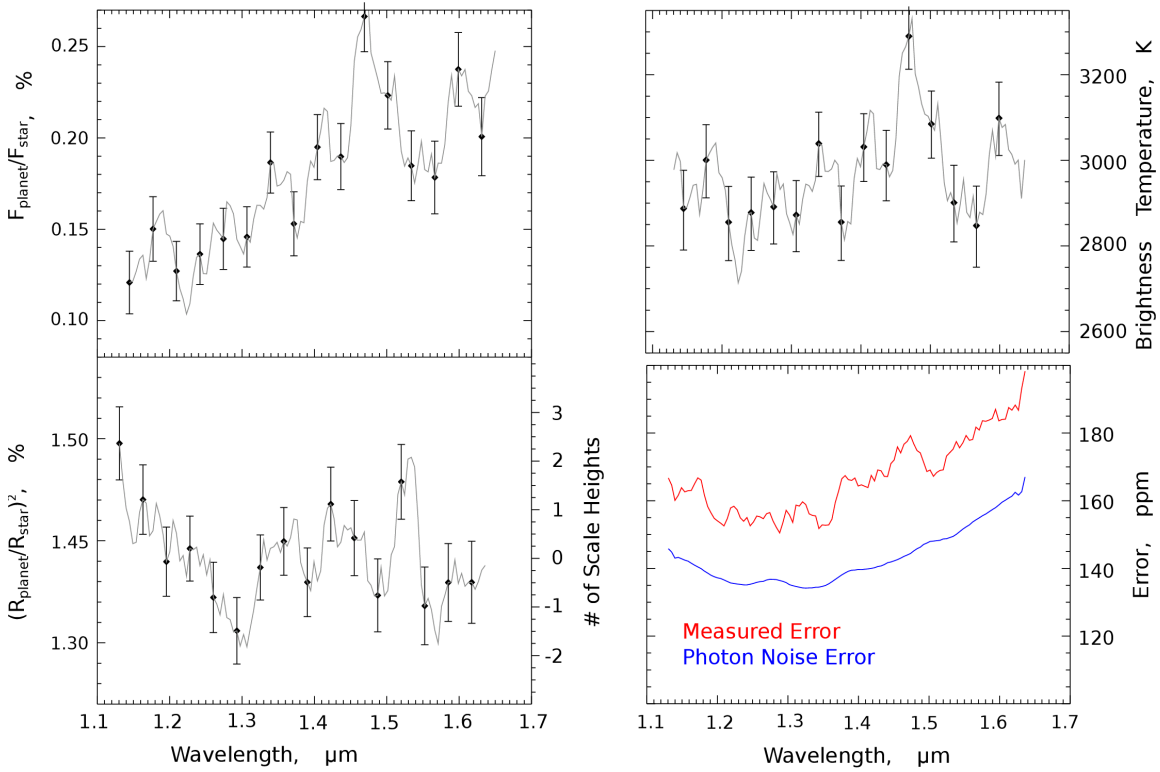

*Figure 10: (Left) Dayside region emission spectrum (top) and terminator region transmission spectrum (bottom). The data averaged in the spectral dimension and statistically independent measurements are shown as solid circles with plus/minus 1-σ errors. The grey line shows the spectrum value computed using a scrolling boxcar. (Right) The dayside emission spectrum in units of brightness temperature (top) and an uncertainty spectrum showing how each channel compares to the photon noise (bottom). These measurements approach the theoretical limit for what is achievable with WFC3 in the IR grism spectroscopy mode.*

For a measurement precision of ~160 ppm, corresponding to averaging 7 pixel-based spectral channels together in these observations, the 256×256 subarray mode delivers nearly ideal noise properties. When the entire 120-pixel passband is averaged together, the measured noise of ~52 ppm is about 1.45 times the theoretical instrument precision (see Figure 9). This is shows that WFC3 IR is a very good instrument for exoplanet characterization. Instrument systematic errors are present, but they average down relatively well. Two separate instrument systematics become detectable in the WASP-12b data when the entire spectral passband is averaged together. The first of these systematics is an intra-orbit, linear flux change that is easily detrended based on the out-of-eclipse data. The second of these systematics is a pattern in the data for each orbit associated with the third (and final) buffer dump; these data have a systematically lower flux value that becomes apparent when averaging the data together for the entire passband. Attempts to decorrelate using the optical state vectors did not work, but an OOT style of approach should prove effective. We did not pursue further correction of this particular systematic because it does not interact with the science related analysis for this paper. Additional observations of both science targets and calibrator stars would benefit our understanding of the ultimate dynamic range capabilities of the instrument.

From the point of view of extracting a spectrum of WASP-12b with spectral resolution of $\Delta\lambda/\lambda = 42$, corresponding to averaging 7 pixel-based spectral channels, the measurement noise budget can be modeled as the root sum of squares of several components. Selecting 1.38 μm for the reference wavelength, we model the uncertainty as 140 ppm due to photon shot noise, 16.5 ppm due to electronics noise, 65 ppm due to a systematic component that is wavelength independent, and 55 ppm due to a systematic that is wavelength dependent and affects the 1.38-1.48 μm channels (see Figure 10). This model is a worst case scenario and assumes the systematic errors have a covariance of zero; if this is not the case, the amplitude of the one or both of the systematic error terms is reduced to accommodate the covariance contribution. Given that (1) a calibration method based on a linear detrending of inter-orbit data results in a measurement error that is within 15% of the theoretical limit, (2) we wish to avoid the debate associated with

259 more complex systematic error removal, and (3) our measurement uncertainties
260 incorporate the affect of the small residual systematic errors, we proceed with the science
261 analysis using the calibration based on the simple linear detrending of the inter-orbit data.

*Table 2:* *The transmission and emission spectrum data determined as described in Section 2.3. The values represent the average of seven pixel-based spectral channels; thus, every seventh wavelength is a statistically independent sample, and there are seven equally valid wavelength grids. The error estimate is derived from a combination of the multiparameter fitting process and the uncertainty in the removal of the Bergfors-6 emission. Data for the wavelength grid we use for plotting the spectra begins with 1.1306 μm and is every seventh row thereafter.*

| Transmission Spectrum | | | Emission Spectrum | | |
|---|---|---|---|---|---|
| [μm] | $(R_p/R_s)^2$ [%] | 1-σ error [%] | [μm] | $F_p/F_s$ [ppm] | 1-σ error [ppm] |
| 1.1306 | 1.498 | 0.018 | 1.1306 | 1355 | 177 |
| 1.1353 | 1.480 | 0.018 | 1.1353 | 1440 | 174 |
| 1.1399 | 1.466 | 0.017 | 1.1399 | 1411 | 170 |
| 1.1445 | 1.459 | 0.017 | 1.1445 | 1208 | 171 |
| 1.1492 | 1.449 | 0.017 | 1.1492 | 1216 | 174 |
| 1.1538 | 1.449 | 0.016 | 1.1538 | 1269 | 172 |
| 1.1584 | 1.464 | 0.017 | 1.1584 | 1336 | 173 |
| 1.1631 | 1.470 | 0.017 | 1.1631 | 1357 | 173 |
| 1.1677 | 1.468 | 0.017 | 1.1677 | 1229 | 175 |
| 1.1723 | 1.452 | 0.018 | 1.1723 | 1335 | 177 |
| 1.1770 | 1.455 | 0.018 | 1.1770 | 1501 | 177 |
| 1.1816 | 1.468 | 0.018 | 1.1816 | 1541 | 172 |
| 1.1863 | 1.462 | 0.018 | 1.1863 | 1582 | 168 |
| 1.1909 | 1.453 | 0.017 | 1.1909 | 1601 | 166 |
| 1.1955 | 1.440 | 0.017 | 1.1955 | 1474 | 164 |
| 1.2002 | 1.445 | 0.017 | 1.2002 | 1462 | 164 |
| 1.2048 | 1.460 | 0.017 | 1.2048 | 1406 | 163 |
| 1.2094 | 1.454 | 0.017 | 1.2094 | 1270 | 162 |
| 1.2141 | 1.440 | 0.016 | 1.2141 | 1175 | 165 |
| 1.2187 | 1.443 | 0.016 | 1.2187 | 1118 | 168 |
| 1.2233 | 1.433 | 0.016 | 1.2233 | 1035 | 169 |
| 1.2280 | 1.446 | 0.016 | 1.2280 | 1091 | 168 |
| 1.2326 | 1.446 | 0.016 | 1.2326 | 1241 | 165 |
| 1.2373 | 1.447 | 0.016 | 1.2373 | 1346 | 164 |
| 1.2419 | 1.438 | 0.016 | 1.2419 | 1363 | 165 |
| 1.2465 | 1.443 | 0.017 | 1.2465 | 1257 | 163 |

*Table 2:* *The transmission and emission spectrum data determined as described in Section 2.3. The values represent the average of seven pixel-based spectral channels; thus, every seventh wavelength is a statistically independent sample, and there are seven equally valid wavelength grids. The error estimate is derived from a combination of the multiparameter fitting process and the uncertainty in the removal of the Bergfors-6 emission. Data for the wavelength grid we use for plotting the spectra begins with 1.1306 μm and is every seventh row thereafter.*

| Transmission Spectrum | | | Emission Spectrum | | |
|---|---|---|---|---|---|
| [μm] | $(R_p/R_s)^2$ [%] | 1-σ error [%] | [μm] | $F_p/F_s$ [ppm] | 1-σ error [ppm] |
| 1.2512 | 1.433 | 0.017 | 1.2512 | 1255 | 164 |
| 1.2558 | 1.441 | 0.017 | 1.2558 | 1390 | 166 |
| 1.2604 | 1.422 | 0.017 | 1.2604 | 1533 | 166 |
| 1.2651 | 1.419 | 0.017 | 1.2651 | 1495 | 165 |
| 1.2697 | 1.414 | 0.017 | 1.2697 | 1472 | 167 |
| 1.2743 | 1.424 | 0.017 | 1.2743 | 1447 | 167 |
| 1.2790 | 1.412 | 0.017 | 1.2790 | 1650 | 165 |
| 1.2836 | 1.411 | 0.017 | 1.2836 | 1629 | 163 |
| 1.2883 | 1.405 | 0.017 | 1.2883 | 1583 | 161 |
| 1.2929 | 1.406 | 0.016 | 1.2929 | 1444 | 164 |
| 1.2975 | 1.398 | 0.016 | 1.2975 | 1406 | 168 |
| 1.3022 | 1.404 | 0.016 | 1.3022 | 1364 | 167 |
| 1.3068 | 1.398 | 0.016 | 1.3068 | 1458 | 164 |
| 1.3114 | 1.407 | 0.016 | 1.3114 | 1428 | 170 |
| 1.3161 | 1.416 | 0.016 | 1.3161 | 1554 | 170 |
| 1.3207 | 1.427 | 0.016 | 1.3207 | 1631 | 171 |
| 1.3253 | 1.437 | 0.016 | 1.3253 | 1630 | 170 |
| 1.3300 | 1.450 | 0.016 | 1.3300 | 1610 | 167 |
| 1.3346 | 1.441 | 0.016 | 1.3346 | 1673 | 167 |
| 1.3393 | 1.446 | 0.016 | 1.3393 | 1865 | 167 |
| 1.3439 | 1.447 | 0.016 | 1.3439 | 1860 | 163 |
| 1.3485 | 1.449 | 0.017 | 1.3485 | 1736 | 164 |
| 1.3532 | 1.444 | 0.017 | 1.3532 | 1743 | 164 |
| 1.3578 | 1.450 | 0.017 | 1.3578 | 1768 | 164 |
| 1.3624 | 1.447 | 0.016 | 1.3624 | 1815 | 166 |
| 1.3671 | 1.452 | 0.016 | 1.3671 | 1800 | 171 |
| 1.3717 | 1.461 | 0.016 | 1.3717 | 1529 | 175 |
| 1.3763 | 1.460 | 0.016 | 1.3763 | 1452 | 179 |
| 1.3810 | 1.439 | 0.017 | 1.3810 | 1542 | 180 |
| 1.3856 | 1.434 | 0.017 | 1.3856 | 1535 | 179 |
| 1.3903 | 1.430 | 0.017 | 1.3903 | 1763 | 179 |
| 1.3949 | 1.426 | 0.017 | 1.3949 | 1873 | 180 |

*Table 2:* *The transmission and emission spectrum data determined as described in Section 2.3. The values represent the average of seven pixel-based spectral channels; thus, every seventh wavelength is a statistically independent sample, and there are seven equally valid wavelength grids. The error estimate is derived from a combination of the multiparameter fitting process and the uncertainty in the removal of the Bergfors-6 emission. Data for the wavelength grid we use for plotting the spectra begins with 1.1306 μm and is every seventh row thereafter.*

| Transmission Spectrum | | | Emission Spectrum | | |
|---|---|---|---|---|---|
| [μm] | $(R_p/R_s)^2$ [%] | 1-σ error [%] | [μm] | $F_p/F_s$ [ppm] | 1-σ error [ppm] |
| 1.3995 | 1.440 | 0.017 | 1.3995 | 1838 | 177 |
| 1.4042 | 1.430 | 0.018 | 1.4042 | 1949 | 178 |
| 1.4088 | 1.435 | 0.018 | 1.4088 | 2037 | 177 |
| 1.4134 | 1.458 | 0.018 | 1.4134 | 2162 | 177 |
| 1.4181 | 1.466 | 0.018 | 1.4181 | 2144 | 181 |
| 1.4227 | 1.468 | 0.018 | 1.4227 | 1871 | 179 |
| 1.4273 | 1.461 | 0.018 | 1.4273 | 1876 | 183 |
| 1.4320 | 1.447 | 0.018 | 1.4320 | 1901 | 182 |
| 1.4366 | 1.449 | 0.018 | 1.4366 | 1897 | 181 |
| 1.4412 | 1.458 | 0.019 | 1.4412 | 1865 | 181 |
| 1.4459 | 1.455 | 0.019 | 1.4459 | 1892 | 185 |
| 1.4505 | 1.456 | 0.019 | 1.4505 | 2068 | 186 |
| 1.4552 | 1.451 | 0.018 | 1.4552 | 2420 | 190 |
| 1.4598 | 1.450 | 0.019 | 1.4598 | 2553 | 189 |
| 1.4644 | 1.450 | 0.018 | 1.4644 | 2593 | 191 |
| 1.4691 | 1.454 | 0.018 | 1.4691 | 2664 | 192 |
| 1.4737 | 1.430 | 0.018 | 1.4737 | 2785 | 194 |
| 1.4783 | 1.428 | 0.018 | 1.4783 | 2464 | 192 |
| 1.4830 | 1.423 | 0.018 | 1.4830 | 2393 | 190 |
| 1.4876 | 1.423 | 0.018 | 1.4876 | 2322 | 189 |
| 1.4922 | 1.437 | 0.018 | 1.4922 | 2244 | 186 |
| 1.4969 | 1.423 | 0.018 | 1.4969 | 2246 | 183 |
| 1.5015 | 1.426 | 0.018 | 1.5015 | 2232 | 184 |
| 1.5062 | 1.443 | 0.018 | 1.5062 | 2206 | 182 |
| 1.5108 | 1.447 | 0.018 | 1.5108 | 2339 | 183 |
| 1.5154 | 1.473 | 0.018 | 1.5154 | 2132 | 184 |
| 1.5201 | 1.479 | 0.018 | 1.5201 | 1927 | 184 |
| 1.5247 | 1.473 | 0.018 | 1.5247 | 1845 | 189 |
| 1.5293 | 1.490 | 0.019 | 1.5293 | 1890 | 190 |
| 1.5340 | 1.491 | 0.019 | 1.5340 | 1847 | 191 |
| 1.5386 | 1.486 | 0.019 | 1.5386 | 1753 | 193 |
| 1.5432 | 1.461 | 0.019 | 1.5432 | 1852 | 192 |

*Table 2:* *The transmission and emission spectrum data determined as described in Section 2.3. The values represent the average of seven pixel-based spectral channels; thus, every seventh wavelength is a statistically independent sample, and there are seven equally valid wavelength grids. The error estimate is derived from a combination of the multiparameter fitting process and the uncertainty in the removal of the Bergfors-6 emission. Data for the wavelength grid we use for plotting the spectra begins with 1.1306 μm and is every seventh row thereafter.*

| Transmission Spectrum | | | Emission Spectrum | | |
|---|---|---|---|---|---|
| [μm] | $(R_p/R_s)^2$ [%] | 1-σ error [%] | [μm] | $F_p/F_s$ [ppm] | 1-σ error [ppm] |
| 1.5479 | 1.428 | 0.019 | 1.5479 | 1981 | 193 |
| 1.5525 | 1.418 | 0.019 | 1.5525 | 1826 | 196 |
| 1.5572 | 1.422 | 0.019 | 1.5572 | 1814 | 194 |
| 1.5618 | 1.416 | 0.019 | 1.5618 | 1911 | 194 |
| 1.5664 | 1.406 | 0.019 | 1.5664 | 1783 | 198 |
| 1.5711 | 1.400 | 0.019 | 1.5711 | 1862 | 198 |
| 1.5757 | 1.418 | 0.019 | 1.5757 | 1861 | 201 |
| 1.5803 | 1.423 | 0.019 | 1.5803 | 1963 | 201 |
| 1.5850 | 1.430 | 0.019 | 1.5850 | 2198 | 201 |
| 1.5896 | 1.430 | 0.019 | 1.5896 | 2342 | 202 |
| 1.5942 | 1.441 | 0.020 | 1.5942 | 2172 | 205 |
| 1.5989 | 1.429 | 0.020 | 1.5989 | 2375 | 201 |
| 1.6035 | 1.434 | 0.020 | 1.6035 | 2340 | 202 |
| 1.6082 | 1.428 | 0.020 | 1.6082 | 2374 | 202 |
| 1.6128 | 1.430 | 0.020 | 1.6128 | 2255 | 206 |
| 1.6174 | 1.430 | 0.020 | 1.6174 | 2231 | 205 |
| 1.6221 | 1.426 | 0.020 | 1.6221 | 2166 | 207 |
| 1.6267 | 1.430 | 0.020 | 1.6267 | 2187 | 205 |
| 1.6313 | 1.436 | 0.020 | 1.6313 | 2006 | 213 |
| 1.6360 | 1.438 | 0.021 | 1.6360 | 2229 | 220 |

262

Determining the depth of the primary and secondary eclipse, as a function of wavelength, provides different physical insights into the planetary atmosphere. For the primary eclipse, the depth is a measurement of $(R_p/R_s)^2$ where $R_p$ is the radius of the planet and $R_s$ is the radius of the parent star. For the secondary eclipse measurement, the depth is the ratio of $F_p/(F_s + F_p) \sim F_p/F_s$, where $F_p$ is the flux density of the primary and $F_s$ is the flux density of the parent star. In addition to working with these parameters, we find it useful to represent the primary eclipse (transmission) spectrum in units of scale height and the secondary eclipse (emission) spectrum in units of brightness temperature (Figure 10); for calculating the brightness temperature, we use a model for WASP-12 (Castelli & Kurucz 2003). To provide context, we compare WASP-12b to the only two other hot-Jupiter class planets with near-infrared emission spectra measured with HST (see Figure 11 and Table 3).

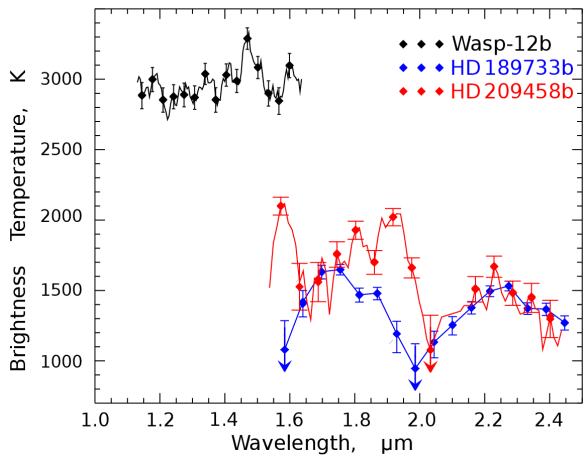

*Table 3:* *Assumptions for this calculation include a Bond albedo = 0.00, Internal energy = 0.00, Molecular weight = 2, and perfect heat redistribution. Differences between the planet temperature estimate in this table and the measured brightness temperature are due to these assumptions.*

| Parameter | WASP-12b | HD 209458b | HD 189733b |
|---|---|---|---|
| Planet Temperature | 2628 K | 1446 K | 1182 K |
| Scale Height (Hs) | 1128 km | 591 km | 249 km |
| 1 Hs precision equivalent | 238 ppm | 182 ppm | 141 ppm |

*Figure 11:* *(Left) Brightness temperature spectra for the three hot-Jovians with emission spectra measured by HST. (Right) Placing WASP-12 in context.*

## 3. Atmospheric Modeling

We explored constraints provided by the spectral time series for the composition of the terminator region atmosphere, the composition and temperature structure of the dayside atmosphere, and the spatial distribution of dayside flux density. Interpreting exoplanet spectra via models is the norm in the field. The approaches we used have been discussed in previous work. Specifically, we use two Bayesian retrieval approaches to analyze both the transmission and emission spectra. The first approach, the optimal estimation approach (Lee et al. 2011, Line et al. 2012), determines the best fit compositions and temperatures by minimizing a quadratic (reduced $\chi^2$) cost function. To

284  determine the uniqueness of the solution, multiple initial composition and temperature
285  guesses are used. To account for non-Gaussian errors (specifically, upper limits), we use
286  the MCMC retrieval approach (Madhusudhan & Seager 2009, Benneke & Seager 2012).
287  The forward model for the emission spectra solves the thermal infrared radiative transfer
288  equation over 90 levels evenly spaced in logP between 30 bars and 1E-6 bars. The
289  transmission forward model solves the limb transmittance radiative transfer equation over
290  45 atmospheric layers evenly spaced in logP between 10 bars and 1E-10 bars. Both
291  models use the HiTemp (Rothman et al. 2010) database for $H_2O$, CO, and $CO_2$, the STDS
292  database (Wenger et al. 1998) for CH4, and the Sharp & Burrows 2007 database for TiO
293  and VO, along with $H_2$-$H_2$/He CIA opacity from Borysow et al. 2001, 2002. We compare
294  emission and transmission models quantitatively using three factors:

1. The reduced $\chi^2$; that is, of $\chi^2$ divided by the total number of data points. Values of 1 suggest that the model fits are generally within 1-sigma of the data errors.
2. The degrees of freedom (DOF), which is the total number of independent pieces of information that can be obtained from the data, as defined by Rodgers 2000. This is not the same as the total number of free parameters. In fact, our definition of DOF often leads to a lesser value than the total number of free parameters, suggesting that not all parameters can be constrained. The DOF is computed from the averaging kernel described in Line et al. 2012.
3. The Bayesian Information Criterion (BIC), which is the $\chi^2 + k \ln(n)$. $k$ is the total number of free parameters and $n$ is the total number of data points. When the total number of parameters increase, a better fit can be obtained, which reduces $\chi^2$; however, the BIC is penalized for those additional parameters. Models with lower BICs are preferred. The addition of a model parameter can only be justified if it improves the fit quality more then the penalization from adding that additional parameter.

## 4. Results

While the primary and secondary eclipse measurements probe separate atmospheric regions and were modeled individually, we used atmospheric "cases" with identical opacity sources for both the primary and secondary eclipse data to provide a uniform basis for comparison. The atmospheric opacity cases span a range from pure $H_2$ continuum opacity to a more physically motivated atmosphere containing $H_2O$, $CH_4$, $CO$, $CO_2$, TiO, and VO. We consider both the WFC3 measurements and Spitzer photometry measurements (Madhusudhan et al. 2011a; Cowen et al. 2012), and the atmospheric models (see Figure 12 and 13) are compared on the basis of goodness of fit and the BIC (see Table 4). Two noteworthy conclusions result from comparing the models. First, on the basis of the existing data, we find no support for including opacity sources beyond $H_2$ continuum. Second, even when we include additional molecular opacity, we find no evidence to support a conclusion of C/O >1. If anything, the modeling suggests that the atmosphere is C depleted.

We stress that these results are indicational due to the limited spectral coverage available. The modeling shows that an improved understanding of the WASP-12b atmosphere requires spectroscopic observations over a larger range of wavelengths. We explored inclusion of TiH and CrH for opacity sources and find that they can be consistent with the data; this exploration was undertaken using the iterative forward model approach with an implementation described by Tinetti et al. (2007a) and

330    subsequently refined in terms of line lists adopted (Tinetti et al. 2007b, Tinetti et al. 2010,
331    Burrows et al. 2005, Burrows et al. 2002). Although TiH and CrH may be consistent with
332    the WFC3 data, we have not undertaken the detailed thermal-chemical modeling that
333    would justify specific selection of these particular metal hydrides given that other
334    possibilities also exist (Sharp & Burrows, 2007). We also note that inclusion of a model
335    for the atmospheric response may be needed to connect the dayside and terminator
336    regions conditions in a detailed way. Despite these limitations, our preliminary modeling
337    is able to provide some insight into the atmosphere of WASP-12b. What emerges is a
338    picture of an exoplanet atmosphere that is like no other exoplanet characterized with
339    infrared spectroscopy to date.

340    **4.1   Terminator Region**
341    The terminator region transmission spectrum of WASP-12b shows maximum opacity
342    at short wavelengths, dropping to a local minimum at 1.3 μm. Additional opacity
343    structure between 1.30 and 1.55 μm is adjacent to the region of minimal opacity between
344    1.56 and 1.64 μm. The opacity slope between 1.1 and 1.3 μm is especially noteworthy.
345    This feature corresponds to about 3 atmospheric scale heights (see Figure 10). While the
346    slope is suggestive of scattering due to dust or aerosols, the slope is too steep to explain
347    in this manner alone. We find that models with $H_2O$, TiO, and VO partially explain this
348    feature, while the $H_2$ continuum and $H_2O/CH_4/CO/CO_2$ models show significant
349    departures from the data in this region (see Figure 12). Although the $H_2$ continuum
350    transmission model has the lowest BIC, the inclusion $H_2O$/TiO/VO is favored over
351    $H_2O/CH_4/CO/CO_2$ based on the combination of $\chi^2$ and BIC values.

## 4.2 Dayside Emission

What immediately stands out in the dayside emission spectrum of WASP-12b is the extremely high brightness temperature of this atmosphere. In the near-infrared, the typical brightness temperature is about 2900 K. The minimum brightness temperature is about 2700 K at 1.22 μm, and the maximum brightness temperature is 3200 K at 1.47 μm. As with the transmission spectrum, an atmospheric-emission model based on $H_2$, with no additional sources of opacity, is preferred by a BIC selection criteria. Unlike the transmission spectrum, if additional opacity sources are added, $H_2O/CH_4/CO/CO_2$ are favored over $H_2O/TiO/VO$ based on a combination of $\chi^2$ and BIC values (see Figure 12). For the $H_2O/CH_4/CO/CO_2$ opacity-based model, we find abundances consistent with an elemental ratio of C/O<1. We also find no evidence for a temperature inversion (see Figure 13). Our result that the model with the most favorable BIC is based on $H_2$ continuum, with no additional opacity, supports the recent findings by Crossfield et al. (2012b), who argue that more complex models are not supported by the data. If we include opacity due to $H_2O/CH_4/CO/CO_2$, our results disagree with the finding of Madhusudhan et al. (2011a) reporting the C/O>1.

*Table 4:* A summary of the opacity sources used to model both the dayside emission and terminator region transmission spectra showing the comparison of the models with the data using the $\chi^2$ and BIC together with retrieved molecular abundance values for each case.

| case | | transmission | | | emission | |
|---|---|---|---|---|---|---|
| Case 1 | C/O | $\chi^2$=2.63 | - | C/O | $\chi^2$=4.24 | - |
| | - | DoF=1.07 | - | - | DoF=1.95 | - |
| | | BIC=14.2 | - | | BIC=19.7 | - |
| Case 2 | C/O | $\chi^2$=2.59 | $H_2O$=1.91E-4 | C/O | $\chi^2$=2.64 | $H_2O$=3.96E-8 |
| | 1.6E-2 | DoF=2.02 | $CH_4$=9.58E-12 | 6.6E-1 | DoF=3.33 | $CH_4$=6.32E-7 |
| | | BIC=25.7 | CO=3.32E-7 | | BIC=30.5 | CO=8.60E-7 |
| | | | $CO_2$=2.90E-6 | | | $CO_2$=2.73E-6 |
| Case 3 | C/O | $\chi^2$=1.50 | $H_2O$=6.32E-5 | C/O | $\chi^2$=4.07 | $H_2O$=7.90E-8 |
| | - | DoF=3.18 | TiO=4.15E-5 | - | DoF=4.32 | TiO=3.23E-6 |
| | | BIC=21.7 | VO=3.97E-5 | | BIC=28.8 | VO=8.79E-6 |
| Case 4 | C/O | $\chi^2$=1.48 | $H_2O$=1.17E-4 | C/O | $\chi^2$=2.36 | $H_2O$=3.44E-8 |
| | 1.3E-3 | | $CH_4$=2.25E-8 | 2.9E-1 | | $CH_4$=8.18E-7 |
| | | DoF=3.37 | CO=2.36E-7 | | DoF=5.38 | CO=7.98E-7 |
| | | | $CO_2$=2.00E-8 | | | $CO_2$=5.78E-6 |
| | | BIC=30.4 | TiO=5.23E-5 | | BIC=36.4 | TiO=1.30E-5 |
| | | | VO=5.07E-5 | | | VO=1.30E-5 |

| **Results of Uncertainty Analysis for Atmospheric Retrievals** | | | | |
|---|---|---|---|---|
| molecule | **Emission Spectra Abundances** | | **Transmission spectra abundances** | |
| | uncertainty | constraint | uncertainty | constraint |
| H2O | 4.65E-14 - 2.8E-02 | 0.04-poor | 1.57E-05 - 8.65E-04 | 0.95—good |
| CH4 | 8.95E-08 - 6.12E-06 | 0.98-good | 3.15E-12 - 1.61E-04 | 0.07 – poor |
| CO | 1.14E-12 - 5.67E-01 | 0.05-poor | 2.47E-11 - 2.26E-03 | 0.01 –poor |
| CO2 | 7.04E-09 - 2.40E-03 | 0.79-limited | 4.81E-12 - 8.34E-05 | 0.18-poor |
| TiO | 2.42E-06 - 6.13E-05 | 0.98-good | 6.43E-06 - 4.25E-04 | 0.95--good |
| VO | 3.08E-06 - 4.96E-05 | 0.99 -good | 5.07E-06 - 5.06E-04 | 0.97--good |

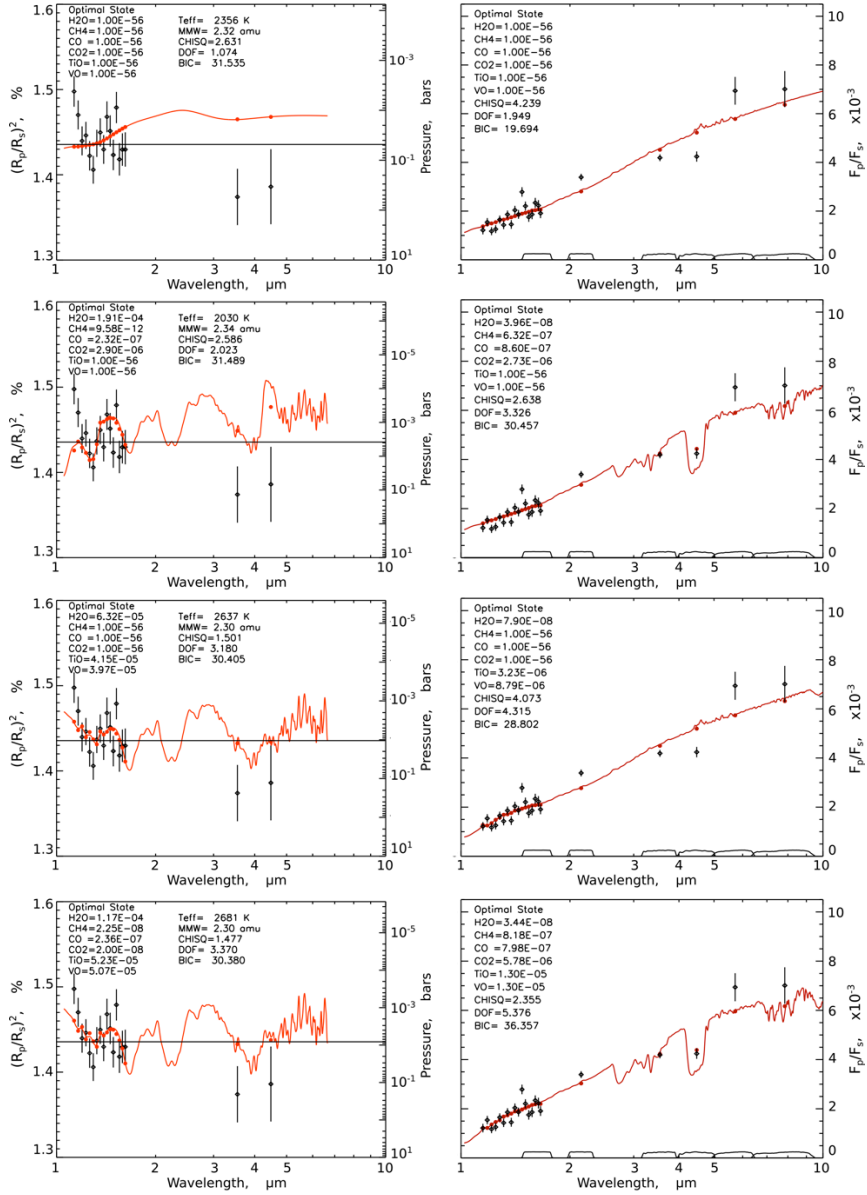

*Figure 12:* Data and best fitting models for the dayside emission (left) and terminator region transmission (right) considering four opacity scenarios listed in Table 4. The model based on $H_2$ with no additional opacity sources has the lowest BIC, implying that more observational data is needed to constrain the presence of additional opacity sources. Molecular abundance and model parameters are listed as inset text for each scenario.

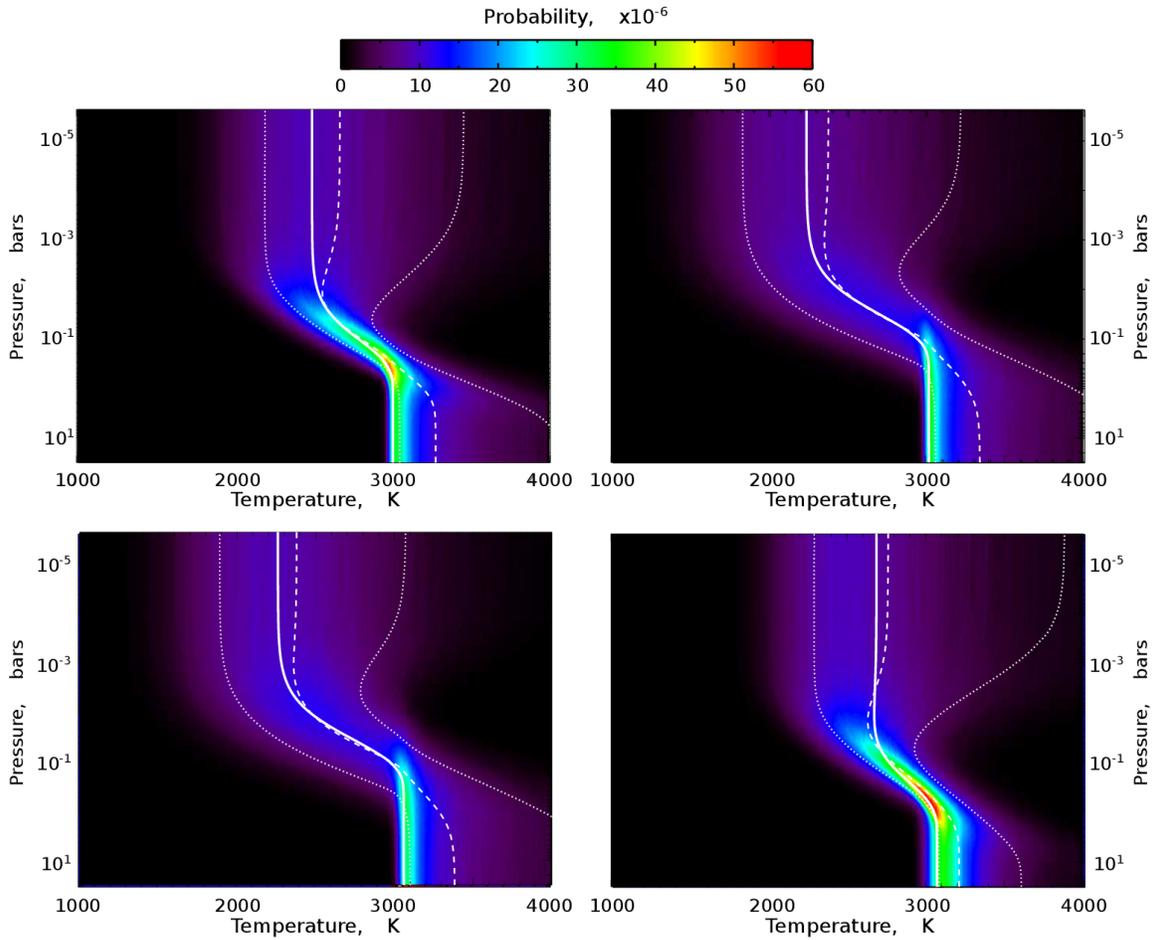

*Figure 13: Temperature profiles retrieved for each of the four dayside atmosphere cases listed in Table 4, with case 1 in the upper left location and moving clockwise for the other cases. The solid line is the "best-fit" temperature profile and the two dotted curves are the +/- 1 sigma. The dashed curve in the middle is the "median" temperature given all of the degeneracies amongst the temperature parameters and gasses.*

368

**Discussion:**

This is the first time that high fidelity, space-based spectroscopy of a planet this hot has been obtained. The spectra of the terminator and dayside regions are separated by 28.4 hours and come as close to taking a spectral "snap-shot" of the planet's atmosphere as is observationally possible with HST. An advantage of temporally closely spaced observations is that they minimize the potential for stellar variations or climate variability in the exoplanet. Although we treated these regions separately in our modeling effort, we were able to find that an $H_2$ atmosphere, with no additional opacity sources, is the optimal using a BIC criterion. If additional opacity source are present, we find no evidence for C/O>1 and, thus, simultaneously confirm the findings of Crossfield et al. (2012b) and differ from the findings of Madhusudhan et al. (2011a). In cases where additional molecular opacity was added, we found that CO, TiO, and VO tended to have similar abundance solutions for emission and transmission spectra, whereas $H_2O$ and $CH_4$ tended to have substantially different dayside and terminator abundances; the behavior of $CO_2$ was in between these two extremes. While the pure $H_2$ atmosphere is preferred in some sense by the data, a more physically plausible atmosphere is likely to contain some molecules. Our modeling shows that the pressure scale probed by these observations is uncertain and will remain so until the atmospheric composition can be better constrained either through additional data or detailed modeling.

The dayside brightness temperature of WASP-12b is sufficiently high that it is the same as the temperature of red dwarf stars of M4V to M5V class and, thus, represents an extreme case of the hot-Jovian class exoplanets. Only two other hot-Jovian type planets have had HST measurements of the near-infrared emission spectrum, and they have significantly lower near-IR brightness temperatures than WASP-12b (see Figure 11). Given the extreme nature of the WASP-12b atmosphere, perhaps it is not surprising that opacity due to TiO, VO and also TiH and CrH may be compatible with the data. Because we do not undertake any thermo-chemical modeling, we consider our models to be preliminary and accept they do not provide a complete view of the planet's atmosphere. Opacity due to TiO/VO has been proposed (Hubeny et al. 2003, Fortney et al. 2008) as a mechanism to produce temperature inversions in hot-Jovian type exoplanets. The currently available data in the near-infrared and mid-infrared, are not sufficient to directly constrain the opacity sources or the C/O ratio. Observations with improved spectral coverage, and higher SNR, would likely provide the needed additional constraints to dramatically improve our ability to understand the WASP-12b atmosphere.

The proposed prolate distortion of WASP-12b (Li et al. 2010) is, in principle, observable in the primary eclipse light curve. The presence of a prolate distortion causes the eclipse depth to be slightly deeper at the edges because the projected area of the planet is increased slightly for angular displacements away from the eclipse center. Using Spitzer primary eclipse measurements, Cowan et al. (2012) found indications of a larger than predicted prolate distortion at measured 4.5 μm. Using our primary eclipse measurements, we searched for the degree of prolate distortion consistent with the data (see Figure 14). Our measurements, at the 95% confidence level, constrain the prolate distortion to be less than 1.34 and are, thus, consistent with the degree of the prolate distortion proposed by Li et al. (2010) and are inconsistent with the large values of prolate distortion found by Cowan et al. (2012). This difference with the Cowen et al. result could be due to probing different regions of the planetary atmosphere.

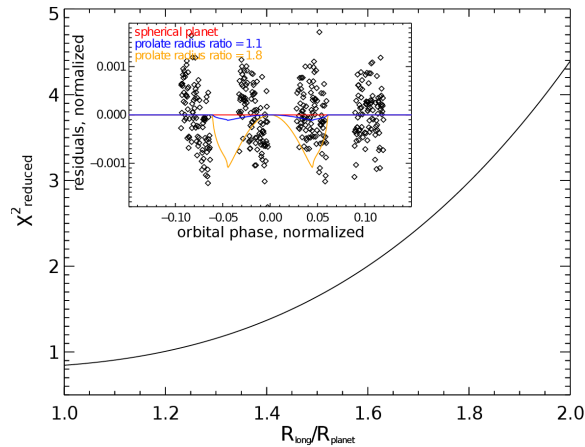

***Figure 14:*** *Results of a $\chi^2$ analysis of the degree of prolate distortion consistent with the WFC3 data. The data are consistent with either a small degree of prolate distortion (including zero) while large degrees of prolate distortion are inconsistent with our measurements. Here, the degree of prolate distortion is measured by the ratio of radii of the planet along the axis pointing to the star, $R_{long}$, and the radii of the planet perpendicular to the star, $R_{planet}$.*

## 5. Conclusions

We present WFC3-IR grism spectroscopy measurements of the dayside and terminator regions of WASP-12b. Our measurements show this is the hottest exoplanet thus characterized to date with a brightness temperature maximum of 3200 K. For both emission and transmission measurements, we find the atmospheric model with the lowest BIC is pure $H_2$ without additional opacity sources. We have explored more physically realistic model atmospheres containing a range of molecules and find no evidence of C/O>1. This finding supports the results of Crossfield et al. 2012b and differs from the results of Madhusudhan et al. (2011a). We also find that TiH and CrH opacity is consistent with the WFC3 spectrum, suggesting that further investigation into sources of opacity in the atmospheres of very hot giant planets is desirable. From the standpoint of temperature and opacity sources, the atmosphere of WASP-12b appears more star like than planet like, confirming that this planet is truly an extreme world. These results, together with previous observational and theoretical work, support the conclusion that hot-Jupiter-type exoplanets are not a uniform class of objects and that these worlds exhibit a diversity that we are only beginning to understand.

From an instrumental point of view, this paper reports the first exoplanet emission spectrum measured with the WFC3 instrument. We also conducted a systematic analysis of 65 orbits of WFC3-IR grism spectroscopy observation to characterize instrument systematics. To our knowledge, the scope and detailed level of this instrument analysis is unique in the area of exoplanet spectroscopy. We find that instrument systematics originating in the detector system can vary considerably. The standard pipeline can mask these problems, and we strongly encourage other researchers to analyze WFC3-IR measurements at the individual non-destructive read (Nsamp) level. Notwithstanding the problems with some detector modes, the WFC3 instrument is an excellent tool for exoplanet spectroscopic characterization; when operated in the optimal detector readout mode, WFC3-IR grism spectroscopy delivers nearly photon-noise-limited measurements with a dynamic range of ~6000:1 that do not require complex decorrelation to correct systematic errors.

**Acknowledgements:**
The research described in this publication was carried out in part at the Jet Propulsion Laboratory, California Institute of Technology, under a contract with the National Aeronautics and Space Administration. Y.F. and H.K are supported by JSPS (Japan Society for the Promotion of Science) Fellowship for Research, DC:23-6070 and

449  PD:22-5467, respectively. We thank Nikku Madhusudhan for kindly providing the data
450  for a previously published theoretical model for comparison with these observations. We
451  are grateful to Rachel Akeson and Thomas Green for useful discussions and suggestions
452  on improving the manuscript. Copyright 2012. All rights reserved.